\documentclass[11pt]{article}
\usepackage{amsmath}
\usepackage{amssymb}
\usepackage{graphicx}
\usepackage{hyperref}

\def\be{\begin{equation}}
\def\ee{\end{equation}}
\def\bea{\begin{eqnarray}}
\def\eea{\end{eqnarray}}
\def\ba{\begin{array}}
\def\ea{\end{array}}

\def\pd{{\partial}}

\def\a{{\alpha}}

\def\m{{\mu}}
\def\n{{\nu}}

\def\pc{{\dot\phi_c}}
\def\cA{{\cal A}}

\title{Backreaction of accreting matter onto a black hole in the
Eddington-Finkelstein coordinates
\date{}}

\author{E. Babichev\,$^{a,b},$ V. Dokuchaev\,$^{b},$ Yu. Eroshenko\,$^{b}$
\vspace{.1cm}\\
\normalsize\llap{$^a$}
 \it Laboratoire de Physique Theorique d'Orsay CNRS UMR 8627, \\
 \normalsize\it 	Universit\'{e} Paris-Sud 11 91405 Orsay Cedex, France,\\
\normalsize\llap{$^b$}
\it Institute for Nuclear Research of the
Russian Academy of Sciences, \\
      \normalsize \it  60th October Anniversary Prospect, 7a, 117312
      Moscow, Russia
\vspace{-.25cm}
}

\begin{document}

\maketitle

\begin{abstract}
We study backreaction of accreting matter onto a spherically symmetric black hole in a perturbative way, when accretion is in a quasi-steady state. General expressions for corrections to the metric coefficients are found in the EddingtonÐFinkelstein coordinates. It is shown that near the horizon of a black hole, independently of the form of the energyÐmomentum tensor, the leading corrections to the metric are of the Vaidya form. The relation to other solutions is discussed and particular examples are presented.
\end{abstract}


\section{Introduction}
Study of spherically-symmetric black hole accretion has a long story. In Newtonian gravity, first the problem of matter accretion onto compact objects has been formulated in a self-similar manner by Bondi \cite{Bondi}. Later, Michel investigated steady-state spherically symmetric flow of a test gas onto a Schwarzschild black hole in the framework of General Relativity  \cite{Michel}. Accretion of a perfect fluid with a general equation of state onto a Schwarzschild black hole has been investigated in \cite{BDE,BDE05}, and a similar analysis for a charged black hole has been done in \cite{BDE2}. In \cite{Petrich} an exact analytic solution of non-spherical accretion of a stiff fluid onto a rotating black hole was derived, and a generalization of this analysis to the case of a rotating charged black hole was done in \cite{BDE3}. Other studies of spherically symmetric accretion of different types of fluids onto black holes and wormholes were undertaken in a number of works \cite{accretion}.
 See also an excellent review on black holes in FRW universe \cite{Carr:2010wk}.

Usually in analytic study of the problem of accretion  an important simplification is made --- backreaction of an accreting matter is neglected. In practice this assumption means that the accreting matter is very light. It is, however, of a great interest to investigate backreaction effects on the metric of a black hole. To solve this problem in a full way is notoriously difficult, only a few analytic solutions  are known, where backreaction is taken into account. In particular,  the celebrated Lema\^{i}tre-Tolman solution describes a spherically symmetric collapsing dust \cite{Lemaitre}. 
Another well-known solution is the Vaidya solution \cite{Vaidya}, which may describe a black hole surrounded by radially moving infalling photons. 
Generalizations of the Vaidya solution to include more general energy-momentum tensor are possible, see e.g. Ref.~\cite{Wang:1998qx}.

In this paper we take a different approach to attack the problem of backreaction of accreting matter.
Instead of looking for particular exact solutions, we implement 
a perturbative scheme: for an accreting matter with general energy-momentum tensor we find corrections to the metric due to backreaction. The scheme goes as follows. Zero approximation is a solution for an accreting matter in the test fluid approximation, i.e. when backreaction of accreting matter is neglected. The first correction to the metric are found from the Einstein equations, where the source term is computed using zeroth-order solution for matter. Although the perturbative approach is not designed to find exact solutions, neither to take into account strong backreaction effects, it allows to study the problem in general manner, without relying on a specific choice of the energy-momentum tensor. We also note, that a similar approach was implemented in \cite{doker11} for calculating backreaction effects using the Schwarzschild (static) coordinates. Here, instead, we use ingoing Eddington-Finkelstein coordinates, in which case we avoid problems related to the singularity of the coordinate system at the horizon of a black hole.

The paper is organized as follows. In Sec.~\ref{Scheme} the general perturbation scheme is described;
in Sec.~\ref{Schwarzschild} corrections to the metric are found for the case of accretion onto a Schwarzschild black hole according to the scheme, 
and in Sec.~\ref{RN} for the case of a Reissner-Nordstr\"om black hole. 
In Sec.~\ref{Perfect}, the results are applied to accretion of a perfect fluid with an arbitrary equation of state. 
The shift of the apparent horizon is discussed in Sec.~\ref{Shift}. In Sec.~\ref{Relation}
 we compare corrections found with the use of the perturbative approach to known exact solutions. 
 It is emphasized that near the horizon corrections to the metric are of the Vaidya form, 
 independently of the energy-momentum tensor of the accreting matter. 
 Sec.~\ref{Examples} contains particular examples:  in Sec.~\ref{Stiff} corrections  are computed for accretion of a stiff fluid; 
while in Sec.~\ref{scalar} for accretion of a canonical scalar field and a galileon field. We conclude in Sec.~\ref{Conclusion}.

\section{Corrections to the metric}
\label{Formalism}

\subsection{Perturbation scheme}
\label{Scheme}
In general case of a black hole accretion the full solution to the Einstein equations are not known. Indeed, there are only a few special cases of the energy-momentum tensor, for which the exact solutions were found, e.~g., the Lema\^{i}tre-Tolman  \cite{Lemaitre} and the Vaidya \cite{Vaidya} solutions. Here we proceed with the perturbative approach to the problem. We find corrections to the metric due to accretion of matter with a general energy-momentum tensor, $T_{\mu\nu}$, in the (almost) steady-state regime of accretion.
In this regime the energy-momentum tensor is a function of the radial coordinate  only. The solution for the energy-momentum tensor $T_{\mu\nu}(r)$ is assumed to be found in the test fluid approximation, i.e. without taking into account backreaction. Then, using $T_{\mu\nu}(r)$ at zeroth-order approximation
we can find first-order corrections to the metric due to the backreaction of matter.

More precisely, our approach is as follows. The full set of equations consist of the Einstein equations,
\be
\label{g eom}
G_{\m\n}\left[ g_{\m\n} \right] = 8\pi T_{\m\n}\left[g_{\m\n},\phi \right],
\ee
and the equations of motion for matter,
\be
\label{eom phi}
\mathcal{E}\left[\phi,g_{\m\n}\right] = 0,
\ee
where by $\phi$ we denote collectively the degrees of freedom for the accreting matter.
For example, in the case of accretion of a single scalar field, $\phi$ is a scalar field itself.
The equations of motion for matter (\ref{eom phi}) can be obtained from the Bianchi identities, therefore they are not independent, but in practice it is convenient to keep them. 

A zeroth-order approximation is found by neglecting backreaction of the matter in Eq.~(\ref{g eom}). Put differently, the solution in this approximation is the vacuum solution for the metric, $g^{(0)}_{\m\n} = g^{vac}_{\m\n}$, so that
\be
\label{g eom 0}
G_{\m\n}\left[ g^{(0)}_{\m\n} \right] = 0.
\ee
In the same approximation, the solution for the matter field(s), $\phi^{(0)}$, is found from,
\be
\label{eom phi 0}
\mathcal{E}\left[\phi^{(0)},g^{(0)}_{\m\n}\right] = 0.
\ee
Solutions at this zeroth-order approximation has been used in literature extensively.

A step beyond the test-fluid approximation is to find corrections to the metric due to backreaction of matter.
In a first-order approximation we can substitute $\phi^{(0)}$ and $g^{(0)}_{\m\n}$ in the r.h.s. of (\ref{g eom}), so that the Einstein equations take the form,
\be
\label{g eom 1}
G_{\m\n}\left[ g^{(0)}_{\m\n} + g^{(1)}_{\m\n} \right] = 8\pi T_{\m\n}\left[g^{(0)}_{\m\n},\phi^{(0)} \right].
\ee
Assuming that $g^{(1)}_{\m\n}$ is small with respect to $g^{(0)}_{\m\n}$ one can linearize the above equation to obtain
a system of linear partial differential equations.  We will implement this approach below.

\subsection{Schwarzschild black hole}
\label{Schwarzschild}

The general spherically symmetric line element can be written in the following form, (see, e.\,g., \cite{Bondi64,LL}):
\begin{equation}
    \label{EF}
    ds^2 = e^{\nu(V,r)+2\lambda(V,r)}dV^2 - 2 e^{\lambda(V,r)}dVdr -r^2d\Omega,
\end{equation}
where $\nu(V,r)$ and $\lambda(V,r)$ are two arbitrary functions.
This is a coordinate frame of the Eddington-Finkelstein (EF) type,
related to the infalling null (photon) geodesics.
The metric (\ref{EF}) is similar to one introduced by Bondi {\it et. al.} \cite{radiationcoordinates} using ``radiation coordinates''.
The vacuum Schwarzschild solution is recovered by setting $\lambda =0$ and $e^{\nu(V,r)} = 1- 2M_0/r $,
where $M_0$ is the mass of the black hole, so that
\begin{equation}
    \label{EF0}
    ds^2_\text{vac} = \left( 1 - \frac{2M_0}{r} \right) \, dV^2 - 2 dVdr -r^2d\Omega,
\end{equation}
Note that in our conventions Eq.(\ref{EF0})
is the solution for the metric at zeroth-order approximation.

Analogously, instead of the metric coefficient $\nu(V,r)$ we will use a function $M(V,r)$, defined as follows,
\begin{equation}
    \label{corr}
    e^{\nu(V,r)}  \equiv 1-\frac{2M(V,r)}{r},
\end{equation}
so that at zeroth-order approximation, $M(V,r) = M_0={\rm const}$.

Plugging (\ref{corr}) and (\ref{EF}) into the perturbed Einstein equations (\ref{g eom 1}) we obtain the following system of equations
(other components of  (\ref{g eom 1}) are trivial),
\begin{eqnarray}
    	8\pi T_0^{\phantom{0}0} &=&  -e^\nu\left(\frac{1}{r^2}+\frac{\nu'}{r}\right)+\frac{1}{r^2}  , \label{EG00}\\
	 8\pi T_0^{\phantom{0}1}  &=& 	\frac{e^\nu}{r}\dot\nu,       \label{EG01}    \\
	8\pi T_1^{\phantom{0}0}	&=&   \frac{2\left(e^{-\lambda}\right)'}{r}, \label{EG10}\\
	8\pi T_1^{\phantom{0}1}   &=& -e^\nu\left(\frac{1}{r^2}+\frac{\nu'}{r}\right)+\frac{1}{r^2} - \frac{2\lambda'}{r}e^\nu,  \label{EG11}\\
	8\pi T_2^{\phantom{0}2} 	&=& 8\pi T_3^{\phantom{0}3} =  -e^\nu\left(\lambda''+\frac{\nu''}{2}\right)    -e^{-\lambda}\dot{\lambda}' \nonumber \\
	& &- e^\nu\left(\lambda'^2+\frac{\nu'^2}{2}+\frac{\lambda'+\nu'}{r} +\frac32\lambda'\nu'\right).  \label{EG22}
\end{eqnarray}
where dots stand for $\partial/\partial V$, and primes denote $\partial/\partial r$.
Note that according to the general scheme, see Eq.~(\ref{g eom 1}),
the l.h.s. of Eqs.~(\ref{EG00}-\ref{EG22}) contain components of the energy-momentum tensor taken at zeroth-order approximation, i.e. on solutions of the matter field(s) when backreaction is neglected. Not all components of the above system of equation are independent ---
using the Bianchi identities it can be shown that  Eq.~(\ref{EG22}) is a combination of  Eqs.~(\ref{EG00})-(\ref{EG11}). 
Thus we will not take Eq.~(\ref{EG22}) into account below.

Substituting (\ref{corr}) into (\ref{EG00}) and (\ref{EG01}) we find,
\begin{equation}
\label{Mprim}
M' = 4\pi T_0^{\phantom{0}0} r^2
\end{equation}
and
\begin{equation}
\label{Mdot}
\dot M =  {\cal A},
\end{equation}
correspondingly, where in the last equation we introduced a notation
\be
\label{flux}
{\cal A} \equiv - 4\pi T_0^{\phantom{0}1} r^2,
\ee
for the total flux of energy crossing some radius $r$. The r.h.s. of Eqs.~(\ref{Mprim}) and (\ref{Mdot}) are taken at zeroth-order approximation.
The components of the energy-momentum tensor are time-independent in the given approximation.
The flux ${\cal A}$ does not depend on $r$ as well, i.e. ${\cal A} =$const. 
For each particular case  of black hole accretion the value of ${\cal A}$ is calculated 
by solving the equations of motion in a fixed metric (normally by finding integrals of motion).
Additional physical requirements can be used to fix solutions.
For example, in the case of a perfect fluid the flux ${\cal A}$ is fixed at the critical point \cite{Michel,BDE}.

Integrating Eqs.~(\ref{Mprim}) and (\ref{Mdot}) we find,
\begin{equation}
\label{MA}
	M(V,r) = C_0 + \cA V+ 4\pi \int_{r_0}^r T_0^{\phantom{0}0}(r)r^2dr,
\end{equation}
where $C_0$ is a constants of integration and $r_0$ is the horizon locus. Similarly, integrating Eq.~(\ref{EG10}), in the limit $\lambda\ll 1$, one obtains
\begin{equation}
\label{lambda}
	\lambda = - 4\pi \int_{r_0}^r T_1^{\phantom{0}0} rdr + s(V),
\end{equation}
where $s(V)$ is an arbitrary function. Note that integration of Eq.~(\ref{EG11}) gives the same result for $\lambda$, Eq.~(\ref{lambda}). In Eq.~(\ref{lambda}) the function $s(V)$ can be set to zero by redefinition of the time coordinate, $e^{s(V)}dV\to dV$. The constant $C_0$, in turn, can be identified with $M_0$ by an appropriate shift $V \to V +$const. Therefore Eqs.~(\ref{MA}) and (\ref{lambda}) can be rewritten as follows,
\bea
\label{master1}
	M(V,r) &=& M_0 + \cA V+ 4\pi \int_{r_0}^r T_0^{\phantom{0}0}(r)r^2dr, \\
\label{master2}
	\lambda(r) &=& - 4\pi \int_{r_0}^r T_1^{\phantom{0}0} rdr.
\eea
The above equations are the central result of this work.

If the components of the energy-momentum tensor are slowly varying functions of the radial coordinate (which is a reasonable assumption for non-pathological matters), then from (\ref{master1}) and (\ref{master2}) one can find the expressions for the corrections of the metric in the vicinity of the black hole horizon,
\bea
\label{m1hor}
	M(V,r) &=& M_0 + \cA V+ 4\pi r_0^2\,  \left(r- r_0\right) T_0^{\phantom{0}0}\mid_{r=r_0}, \\
\label{m2hor}
	\lambda(r) &=& - 4\pi r_0 \left(r- r_0\right) T_1^{\phantom{0}0}\mid_{r=r_0} .
\eea

According to our perturbative scheme, the above equations (\ref{master1}) and (\ref{master2}) are only valid if the corrections are small. Thus we require,
\be
	\label{conditions}
	\left| \cA V \right|  \ll  M_0, \;  \left| 4\pi \int_{r_0}^r T_0^{\phantom{0}0}(r)r^2dr \right|  \ll  M_0,  \;
	\left|  4\pi \int_{r_0}^r T_1^{\phantom{0}0} rdr \right| \ll 1.
\ee
The first of the inequalities in (\ref{conditions}) is simply a condition that the total flux of the infalling matter is smaller than the bare black hole mass. Similarly, the second condition says that the total mass of the matter inside a radius $r$ is smaller than the bare mass of the black hole. The last condition in (\ref{conditions}) has no obvious interpretation, it reflect the fact that the correction to the non-diagonal part of the metric due to the presence of infalling matter must not be too big.

\subsection{Reissner-Nordstr\"om black hole}\label{RN}
We can generalize our results to the case of the Reissner-Nordstr\"om black hole. Accordingly, instead of (\ref{corr}) we now write the metric coefficient $\nu$ in the following form,
\begin{equation}
    \label{nuRN}
    e^{\nu(V,r)} = 1-\frac{2M(V,r)}{r} + \frac{Q^2}{r^2},
\end{equation}
where $Q$ is a charge of a black hole. We assume that the accreting fluid carries no electric charge, otherwise the total charge of the black hole needs to be modified as well. Substituting (\ref{nuRN}) into (\ref{EG00})-(\ref{EG11}) and taking into account a contribution of the electromagnetic field in the total energy-momentum tensor,
\begin{equation}
T_0^{\phantom{0}0}\to T_0^{\phantom{0}0}+\frac{Q^2}{8\pi r^4},~~~ T_1^{\phantom{0}1}\to T_1^{\phantom{0}1}+\frac{Q^2}{8\pi r^4},
\nonumber
\end{equation}
\begin{equation}
T_2^{\phantom{0}2}\to T_2^{\phantom{0}2}-\frac{Q^2}{8\pi r^4},~~~  T_3^{\phantom{0}3}\to T_3^{\phantom{0}3}-\frac{Q^2}{8\pi r^4},
    \label{telm}
\end{equation}
after lengthy but straightforward calculations one can check that the expressions for the metric coefficients $M(V,r)$ and $\lambda(V,r)$ are the same as in the case of the Schwarzschild black hole. So that the results are given by Eqs.~(\ref{master1}) and (\ref{master2}), 
where $r_0$ has now the meaning of the event horizon of a bare charged black hole.

\subsection{Accretion of a perfect fluid}\label{Perfect}

In this section we consider accretion of a perfect fluid onto a black hole. The stress tensor of a
perfect fluid is given by,
\be
\label{perfect}
T_{\mu\nu} = (\rho+p) u_\mu u_\nu - pg_{\mu\nu},
\ee
where $\rho$ is the energy density and $p$ is the pressure of the fluid in the comoving coordinates, and $u^{\mu}$ is a 4-velocity of the fluid. 
In the EF coordinates the 4-velocity is given by
\begin{equation}
    \label{u}
    u^\mu = \left(\frac{1}{\sqrt{f_0+u^2}+u},\; -u,\; 0,\; 0\right),
\end{equation}
where $u\equiv |dr/ds| >0$ is the absolute value of the radial component of 4-velocity in the static coordinates and $f_0\equiv 1-2 M_0/r$.
It is easy to check that all the components of $u^\mu$ and $u_\mu$ are non-divergent at the horizon.
The relevant components of the energy-momentum tensor are
\be
T_0^{\phantom{0}0} =  \frac{\rho\sqrt{f_0+u^2}-pu}{\sqrt{f_0+u^2}+u} ,\label{EMT00} \quad
T_1^{\phantom{0}0} = -\frac{\rho+p}{\left(\sqrt{f_0 + u^2}+u\right)^2}. 
\ee
In the vicinity of the horizon, $f_0\to 0$, we find,
\be
    T_0^{\phantom{0}0}  \to \frac{1}{2}(\rho-p),\quad
    T_1^{\phantom{0}0} \to -\frac{\rho+p}{4u^2}.
\ee
Thus, from (\ref{m1hor}) and (\ref{m2hor}) we can write for the corrections,
\bea
M(V,r) &\approx& M_0+ \cA V+ 2\pi r_0^2 \left(\rho - p\right)\left( r-r_0 \right),\label{midealfl} \\
\lambda (r)&\approx&   \pi r_0 \frac{\rho+p}{u^2}\left( r-r_0 \right).
\eea
The above expressions are valid for any perfect fluid in the vicinity of the horizon of a black hole. Note that the energy flux onto the black hole is given by  (\ref{flux}), with $T_0^{\phantom{0}1}=-(\rho+p)u\sqrt{f_0+u^2}$ for a perfect fluid, see \cite{BDE}. Therefore, as it can be seen from the (\ref{midealfl}), the accretion of phantom energy, $\rho+p<0$, decreases the black hole mass.
Thus we confirm that the backreaction does not affect the results,
which were reported previously in \cite{BDE}, where the zeroth-order approximation has been used.

Note that the expression (\ref{midealfl}) implies, in particular, that light observers, having different $V=const$,
see the black hole with the different masses $M$ (at different stages of accretion).

\subsection{Shift of the apparent horizon}\label{Shift}

The position of the apparent horizon depends on a choice of a coordinate system. For the metric (\ref{EF}) it can be shown that the location of the  apparent horizon, $r_h$, can be found from the following equation \cite{Nielsen:2010gm},
\begin{equation}
    \label{corr0}
	e^{\nu(V,r)} = 0.
\end{equation}
The above expression can be obtained by  requiring that for a radial outgoing photon $dr/dV =0$. Indeed, from $ds^2=0$ one can find the two radial light geodesics,
\begin{equation}
dV=0, \quad
\frac{dr}{dV}e^{-\nu-\lambda}=\frac{1}{2}.
\label{lgeo2}
\end{equation}
The apparent horizon satisfies the condition that the photons do not cross the $r=const$ surfaces in the growing $r$-direction. It gives $dr/dV=0$, and from (\ref{lgeo2}) one obtains $e^{\nu+\lambda}=0$ and hence $e^{\nu(V,r)}=0$. The later is due to regularity of $\lambda$. 
Now, for a Schwarzschild black hole from (\ref{corr0}) and (\ref{master1}) we find,
\be
\label{corr2}
 M_0 + \cA V+ 4\pi \int_{r_0}^{r_h} T_0^{\phantom{0}0}(r)r^2dr  = \frac{r_h}{2},
\ee
which is an implicit equation for $r_h$. It is not difficult to check that for small shifts of the apparent horizon the last term in the l.h.s. of (\ref{corr2}) can be neglected, since it is of the next order in the expansion, so that we find,
\be
r_h \approx 2M_0 + 2 {\cal A} V.
\ee
Thus the leading term in the apparent horizon shift only depends on the total flux $\cA$ and is independent on other components of the energy-momentum tensor.

Similarly, for the shift of the apparent horizon in the case of a charged black hole one has, instead of (\ref{corr2}),
\be
\label{corr3}
 M_0 + \cA V+ 4\pi \int_{r_0}^{r_h} T_0^{\phantom{0}0}(r)r^2dr  = \frac{r_h}{2} + \frac{Q^2}{2 r_h}.
\ee
Again, neglecting the last term in the l.h.s. of the above equation, we obtain, for the case of the Reissner-Nordstr\"om black hole,
\be
\label{rhRN}
r_h \approx M_0 + {\cal A} V +\sqrt{M_0^2 - Q^2 + 2 M_0 \cA V}.
\ee
Note that for the extreme black hole, $M_0 = Q$, the existence of the apparent horizon depends on the sign of the flux, $\cA$. For phantom accretion, the apparent horizon ceases to exist for positive $V$, which means that our perturbation scheme breaks down for this particular case.
The reason for the breaking down of our scheme is that accretion of phantom decreases the black hole mass, therefore any amount of phantom matter converts the extreme black hole to a naked singularity, thus going beyond our quasi steady-state approximation. 
On the other hand, if normal (non-phantom) matter accretes, then (\ref{rhRN}) can be perfectly applied. 
It is worth to mention, that the in the static Schwarzschild coordinates 
the test fluid approximation is violated for accretion of any type (phantom and non-phantom) matter onto the extreme black hole, see Ref.~\cite{doker11}.

\section{Relation to other solutions}
\label{Relation}
Having found the general form of the corrections to the metric due to backreaction, we can now compare our results to other solutions.
In particular, it is interesting to see how our result  is related to known full analytic solutions.

Let us first consider the case when matter is the cosmological constant term.
In this case it is easy to verify that $T_0^{\phantom{0}0} = \rho_\Lambda $,  $T_1^{\phantom{0}0} = 0$ and $\cA = 0$, where $\rho_\Lambda$ is the energy density of the vacuum term. The general expression (\ref{master1}) and (\ref{master2}) give then,
\be
\label{cc}
	M(V,r) = M_0 +  \frac{4\pi}{3} \rho_\Lambda \left( r^3 - r_0^3 \right), \quad	\lambda =0,
\ee
which, after the redefinition $M_0 \to M_0 + 4\pi \rho_\Lambda  r_0^3/3$, can be identified with the Schwarzschild-de Sitter solution.

As a second example, we consider the case of radially infalling photons, so that the energy-momentum tensor is of the form,
\be
T_\mu^{\phantom{\mu}\nu} = -\frac{\dot M}{4\pi r^2} k_\mu k^\nu.
\ee
The only non-zero component of the  energy-momentum tensor which contributes to the corrections of the metric (\ref{master1}) and (\ref{master2})
is $T_1^{\phantom{0}0}$ corresponding to a non-zero total flux, $\cA\neq 0$, while $T_0^{\phantom{0}0}=T_1^{\phantom{0}0} =0$. Thus,
\be
\label{Vaidya}
	M(V,r) = M_0 +  \cA V, \quad	\lambda =0,
\ee
which corresponds to the Vaidya solution \cite{Vaidya} for the constant flux, $\cA=$const.

It is also interesting to observe that in the vicinity of the horizon of a Schwarzschild black hole, independently of the form of the energy-momentum tensor of the accreting fluid, the corrections to the metric are of the Vaidya form. Indeed, for $r=2M_0$, from (\ref{master1}) and (\ref{master2}), we infer that  $M(V,r=r_0) = \cA V$, $\lambda = 0$, which coincides with (\ref{Vaidya}). This means, in particular, that the first correction to the motion of free particles in the vicinity of the horizon does not depend on the energy-momentum tensor of the accreting fluid: it depends on the total flux $\cA$ only. One should note, however, that although the metric itself is asymptotically Vaidya near the horizon, the derivatives of the metric are not the same (not even asymptotically) as in the Vaidya solution. 
Said differently, the Riemann tensor depends (even near the black hole horizon) on type of the accreting matter.

\section{Examples}\label{Examples}
\subsection{Stiff fluid}\label{Stiff}

At zeroth-order approximation the steady-state accretion in the Schwarzschild coordinates
corresponds to the steady-state solution in the EF coordinates.
Therefore we obtain the same integrals of motion (as a functions of $r$ only), the same expression for the critical points,
and finally the same profile of density and velocity as in both coordinate systems.
Thus we can take results of \cite{BDE}, derived in the static coordinate system and apply them here. In particular, for the linear equation of state $p=\rho-\rho_0$  one has
\begin{equation}
 \label{sol2}
 \rho=\frac{\rho_0}{4}+\left(\rho_{\infty}-\frac{\rho_0}{4}\right)
  \left(1+\frac{r_0}{r}\right)\left(1+\frac{r_0}{r^2}\right),
\end{equation}
where $\rho_\infty$ is the energy density at the infinity. The corresponding profile for the velocity $u=u(r)$ can be calculated from the energy flux conservation equation. In the case of a stiff perfect fluid ($\rho_0=0$), we find, see \cite{BDE}, $\rho u^2=r_0\rho_{\infty}/r^4$. The relevant components of the energy-momentum tensor in this case are given by,
\be
T_0^{\phantom{0}0} =  \rho_{\infty}\left(1+\frac{r_0}{r}\right)\left(1-\frac{r_0^2}{r^2}\right),\quad
T_1^{\phantom{0}0}=-2\rho_{\infty}\left(1+\frac{r_0}{r}\right)^2.
\ee
and the flux,
\be
\cA=8\pi\rho_{\infty}r_0^2.
\ee
Therefore the corrections in the vicinity of the horizon take the form
\bea
M &=& M_0+8\pi\rho_{\infty}r_0^2V+4\pi\rho_{\infty}r_0^3 \left(\frac{ x^3}{3}+\frac{ x^2}{2}-x^2 +\frac{1}{6}- \log x\right),\label{Mstiff}\\
\lambda &=& 4\pi\rho_{\infty} r_0^2 \left(x^2+4x- 5 +2  \log{x}\right) \label{lambdastiff},
\eea
where we have introduced a short-hand notation $x\equiv r/r_0$.

\subsection{Scalar field}\label{scalar}
%
Here we consider accretion of a scalar field and find corrections to the metric. The boundary conditions at the infinity are posed by the cosmological evolution,
\be
\dot\phi|_\infty = \dot\phi_c.
\ee
Let us first consider the standard kinetic term (which in the hydrodynamical description correspond to a stiff perfect fluid), so that the action is given by
\be
S_{\rm can} = \int d^4 x \sqrt{-g} \left(\frac12\pd_\mu\phi \pd^\mu\phi\right)
\ee
Solving the Klein-Gordon equation, $\Box\phi =0$, in the metric of the Schwarzschild black hole one can deduce that the stationary non-singular solution for $\phi$ is given by,
\be
\label{cansol}
\phi = \pc\left( V - r- r_0 \log \left(\frac{r}{r_0}\right) \right).
\ee
Taking into account that the energy-momentum tensor for a canonical scalar field is
\be
T_{\mu\nu}  = \pd_\mu\phi \pd_\nu\phi - \frac12 \left(\pd\phi\right)^2 g_{\mu\nu},
\ee
one can easily calculate the relevant components of the energy momentum tensor on the solution (\ref{cansol}),
\be
    T_0^{\phantom{0}0} = \frac{\pc^2}{2}\left(1-\frac{r_0^2}{r^2}\right)\left(1+\frac{r_0}{r}\right),\quad
    T_1^{\phantom{0}0}  = -\pc^2\left(1+\frac{r_0}{r}\right)^2.
\ee
and
\be
\cA = 4\pi \pc^2 r_0^2.
\ee
Substituting the above expressions to (\ref{master1}) and (\ref{master2}) one can see that the correction to the metric are given by (\ref{Mstiff}) and (\ref{lambdastiff}) after identification $\dot\phi_c^2 \to 2 \rho_\infty$, as one could expect.

It is worth to mention phantom accretion in the context of backreation effects.
One of examples for phantom, which was considered in \cite{BDE}, consisted of a scalar field with the ``wrong'' sign in the action. In terms of a perfect fluid this simply means that the density and the pressure of the effective perfect fluid becomes negative. This, in particular, implies, from (\ref{Mstiff}) that the black hole mass decreases during accretion of a phantom field. This result takes into account backreaction of the accreting field and it confirms the conclusion of \cite{BDE}, where the backreaction was neglected.

Now let us consider accretion of a galileon scalar field. The general galileon model consists of five non-trivial terms: the first one is a potential of the form const$\times\phi$, the second one is the canonical scalar field (which we already considered  above) and three other non-linear terms. Here we consider accretion of the galileon  with the action,
\be
S_{\rm gal} = - \int d^4 x \sqrt{-g}\, \left(\pd_\mu\phi \pd^\mu\phi\right) \Box\phi.
\ee
The stress-tensor for this action is,
\be
\label{Tmngal}
T_{\m\n}^{\rm gal}= - 2\phi_{,\mu} \phi_{,\nu}\Box\phi + 2\phi_{,(\mu}\nabla_{\n)} \left(\pd\phi\right)^2
- g_{\m\n}\phi^{,\a} \nabla_\a \left(\pd\phi\right)^2,
\ee
The solution for steady-state spherically symmetric accretion onto a black hole was found in \cite{Babichev:2010kj},
\be
\label{galsol}
\phi = \pc\left( V - r +2 \sqrt{r_0 r} -2 \sqrt{r_0} \log \left(\sqrt{\frac{r}{r_0}}+1\right) \right).
\ee
Substituting (\ref{galsol}) into (\ref{Tmngal}) and (\ref{flux}) one finds for the galileon,
\be
T_0^0 = \pc^2 \frac{3}{x\left( 1+\sqrt{x} \right)} ,\quad
T_1^0 = -\pc^2 \frac{3}{\sqrt{x}\left( 1+\sqrt{x} \right)^2}, \quad
\cA = 12\pi \pc^2 r_0^2.
\ee
Thus the correction to the metric are
\bea
M &=& M_0 + 12\pi \pc^2 r_0^2 V  \nonumber \\
&+&  4\pi \pc^2 r_0^3 \left( 6\left(\sqrt{x}-1\right)
- 3\left(x-1\right)+2\left(x^{3/2}-1\right)
-6\log\left(\sqrt{x}+1\right) \right), \label{Mgal}\\
\lambda & = & 12\pi \pc^2 r_0^2
\left( -\frac{2}{1+\sqrt{x}} + 2\sqrt{x} -4 \log (1+\sqrt{x}) -1 +4\log 2 \right).\label{lambdagal}
\eea

\section{Conclusion}\label{Conclusion}
In the studies of accretion of matter onto black holes, there are basically two analytic approaches. One is to neglect backreaction of an accreting matter onto the metric of the black hole --- which is reasonable when the matter is light --- and to study dynamics of matter on fixed background metric. 
Another way is to try to find solutions for the full system of the Einstein and matter equations. 
There are only a few such solutions known, the reason is that it is in general very hard to find solutions to the full non-linear system of equations.

In this work we studied the backreaction of an accreting fluid on the black hole metric in a perturbative way. Accretion is spherically symmetric and quasi steady-state. In this approximation we assumed that the accretion flow is small and matter is light, so we could use a perturbation scheme. At zeroth-order approximation the accreting matter does not backreact, so this step is equivalent to the test-fluid approximation, Eqs.~(\ref{g eom 0}) and (\ref{eom phi 0}). Then the first-order corrections to the metric are found from the perturbed Einstein equations, in which the source term is given by the accreting matter at zeroth-order approximation, Eq.~(\ref{g eom 1}). In particular, for the case of spherically symmetric accretion onto a Schwarzschild black hole, the resulting equations for the correction are written down in (\ref{EG00})-(\ref{EG22}). The solution for the corrections to the metric coefficients are found in general form, Eq.~(\ref{master1}) and (\ref{master2}), this is the central result of our paper. 
In a similar manner, the general perturbation scheme is applied to the Reissner-Nordstr\"{o}m black hole, and we have found the same form for the correction of the metric, Eq.~(\ref{master1}) and (\ref{master2}).

We compared our perturbative results with known analytic solutions. In particular, we verified that the our scheme gives correct results in the case of the cosmological vacuum term (up to a redefinition of the central mass), as well as for the radially infalling photons, in this case we reproduced the Vaidya solution.

An interesting consequence of the general result is universality of the corrections to the metric in the vicinity of the horizon of a black hole. We found that, if the accretion rate is nonzero, then, independently on the energy-momentum tensor of an accreting fluid, the leading corrections to the metric are of the Vaidya type, Eq.~(\ref{Vaidya}).

We applied the obtained results for calculation of correction to the metric for a generic perfect fluids and also for particular examples: accretion of a stiff perfect fluid
Eqs.~(\ref{Mstiff}), (\ref{lambdastiff}) as well as accretion of a canonical scalar field and the galileon scalar field, Eqs~(\ref{Mgal}), (\ref{lambdagal}).

\paragraph*{Acknowledgments}
The work of V.~D. and Yu.~E. was supported in part by the Russian Foundation for Basic Research grant 10-02-00635 and by the grant of the Russian Leading scientific schools 3517.2010.2.

\end{document}